\documentclass{nature}

\usepackage{bm}
\usepackage{graphicx}

\newcommand{\beeq}[1] {\begin{equation}#1\end{equation}}

\title{Improving randomness characterization through Bayesian model selection}

\author{Rafael D\'iaz Hern\'andez Rojas$^{1}$, Aldo Sol\'is$^2$, Al\'i M. Angulo Mart\'inez$^{2}$, Alfred B. U'Ren$^{2}$, Jorge G. Hirsch$^{2}$, Matteo Marsili$^{3}$ \& Isaac P\'erez Castillo$^{1,4,*}$}

\begin{document}

\maketitle

\begin{affiliations}
 \item Instituto de F\'isica, Universidad Nacional Aut\'onoma de M\'exico. Apdo. Postal 20-364, Cd. Mx., Mexico, C.P. 04510
 \item Instituto de Ciencias Nucleares, Universidad Nacional Aut\'onoma de M\'exico, Apdo. Postal 70-543, Cd. Mx., Mexico, C.P. 04510
 \item The Abdus Salam International Centre for Theoretical Physics, Strada Costiera 11, 34151 Trieste, Italy
 \item London Mathematical Laboratory, 14 Buckingham Street, London WC2N 6DF, United Kingdom
\end{affiliations}

\begin{abstract}
Random number generation plays an essential role in technology with important applications in areas ranging from cryptography to Monte Carlo methods, and other probabilistic algorithms. All such applications require high-quality sources of random numbers, yet effective methods for assessing whether a source produce truly random sequences are still missing. Current methods either do not rely on a formal description of randomness (NIST test suite) on the one hand, or are inapplicable in principle (the characterization derived from the Algorithmic Theory of Information), on the other, for they require testing all the possible computer programs that could produce the sequence to be analysed. Here we present a rigorous method that overcomes these problems based on Bayesian model selection. We derive analytic expressions for a model's likelihood which is then used to compute its posterior distribution. Our method proves to be more rigorous than NIST's suite and Borel-Normality criterion and its implementation is straightforward. We applied our method to an experimental device based on the process of spontaneous parametric downconversion to confirm it behaves as a genuine quantum random number generator. As our approach relies on  Bayesian inference our scheme transcends individual sequence analysis, leading to a characterization of the source itself.
\end{abstract}

Random numbers have acquired an essential role in our daily lives because of our close relationship with communication devices and technology. There are also numerous scientific techniques and applications that rely fundamentally on our ability for generating such numbers and typically pseudo-random number generators (pRNGs) suffice for those purposes. A new alternative has been proposed by exploiting  the inherently probabilistic nature of quantum mechanical systems. These Quantum Random Number Generators (QRNGs) are in principle superior to their classical counterparts and recent experiments have shown\cite{Solis2015} that they can reach the same quality as commercial pRNGs. However, the natural question of how to assess whether a sequence is truly random is not yet fully established. Pragmatically, the NIST test suite\cite{NIST} has become the standard method for analysing sequences coming from a RNG. The suite is based on testing certain features of random sequences that are hard to reproduce algorithmically, such as its power spectrum, longest string of consecutive \texttt{1}'s, and so on. Even though it constitutes an easily applicable procedure, recent findings show that its reliance on $P$-values is a drawback\cite{second-level-nist, asa-pvalues}, while its lack of formality is a major disadvantage. On the other hand, although no definition of randomness is deemed absolute, a rigorous characterization is presented by the Algorithmic Theory of Information (ATI) but it is unfortunately inapplicable in real cases\cite{Calude2010b}. An alternative which overcomes both formal and applicability issues is the Borel-normality criterion\cite{Calude1993} (BN).  Intuitively, this approach works by successively compressing a given dataset, e.g. $\hat{s}=\{\texttt{0101010010101010101011010}\cdots\}$ of $M$ bits, by taking strings of $\beta$ consecutive bits and computing the frequency of occurrences $\gamma_i^{(\beta)}$ of each of those $i=0,1,\ldots, 2^{\beta}-1$ possible strings.
For example, $\beta=1$ corresponds to looking for the frequencies of the strings $\{\texttt{0},\texttt{1}\}$ in the dataset $\hat{s}$, while $\beta=2$  corresponds to analysing the frequencies of the strings  $\{\texttt{00},\texttt{01},\texttt{10},\texttt{11}\}$, and so on. The whole sequence is said to be Borel-normal if the frequencies are bounded individually according to
\begin{equation}
\left| \gamma^{(\beta)}_i - \frac{1}{2^\beta} \right| < \sqrt{\frac{\log_2 M}{M}},
 \label{eq:borels bound}
\end{equation}
and with $\beta$ an integer ranging from $1$ to $\beta_{{\rm max}}=\log_2\log_2 M$.  It is important to mention that BN criterion is a (nearly) necessary condition for a sequence to be considered random\cite{Calude2010b}. Note that this test is restricted to a-single-sequence classification, so it cannot determine the random character of the generating \emph{source}. 

In the present work, we show that randomness characterization can also be addressed using a Bayesian inference approach for model selection\cite{Haimovici2015},  borrowing the compression scheme of BN. For simplicity, for a fixed $\beta$ we denote each string with its decimal base representation $j\in\{0,1,\ldots,2^{\beta}-1\}\equiv \Xi_\beta$. The first step consists in identifying the models which could have generated a compressed dataset $\hat{s}$. For instance if $\beta=1$, we can describe it as $M$ realizations of a Bernoulli process, leading to two possible models: with and without bias. Similarly, for $\beta=2$, a model represents a way of constructing $\hat{s}$ with bias in some of the $2^2$ possible strings. A simple combinatorial counting reveals that all the possible bias assignations correspond to all partitions of the four strings of $\Xi_2$. 

Thus, in general, given the set  $\Xi_\beta$, let $\mathcal{P}_{\Xi_\beta}$ denote the family of its $B_{2^\beta}=\sum_{K=1}^{2^\beta} \left\{ {2^{\beta} \atop K} \right\}$ possible partitions\cite{partitions}, with $B_{2^\beta}$ the Bell's numbers and $\left\{ {2^\beta \atop K}\right\}$  the Stirling numbers of the second kind, which counts the different ways of grouping $2^\beta$ elements into $K$ sets. Formally, $\alpha^{(K)}_{\ell}=\{\omega^{(1)}_{\ell},\ldots,\omega^{(K)}_{\ell}\}\in\mathcal{P}_{\Xi_{\beta}}$ would refer to the $\ell$-th partition into $K$ subsets, but for notational simplicity we will omit henceforth the index $\ell$. To each partition $\alpha^{(K)}$ there corresponds a unique model $\mathcal{M}_{\alpha^{(K)}}$ which assigns a probability $p_j$ to string $j\in\Xi_{\beta}$ according to the following rule:
\beeq{
\mathcal{M}_{\alpha^{(K)}}=\left\{p_j=\frac{\theta_r}{|\omega^{(r)}|}; \quad \forall r=1,\ldots, K ;\ \forall j\in \omega^{(r)}   \right\}\,.
}
This means that all strings contained in a given subset $\omega^{(r)}$ are deemed equiprobable within the specified model. Thus, keeping $\beta$ fixed, the likelihood of observing the given dataset $\hat{s}$ in a model $\mathcal{M}_{\alpha^{(K)}}$ is:
\beeq{
P\left(\hat{s}\big|\mathcal{M}_{\alpha^{(K)}},\{\theta_r\}_{r=1}^K\right)=\prod_{r=1}^K\left(\frac{\theta_r}{|\omega^{(r)}|} \right)^{k_{\omega^{(r)}}}\,,
}
where $k^{(\beta)}_j$ is the frequency of string $j\in\Xi_{\beta}$ and we have defined $k_{\omega^{(r)}}=\sum_{j\in\omega^{(r)}}k^{(\beta)}_j$ as the aggregate frequencies of the strings in the subset $\omega^{(r)}$. (For further use, we also introduce the relative aggregate frequencies $\gamma_{\omega^{(r)}}=\frac{\beta}{M}k_{\omega^{(r)}}$.) From this perspective, only the model that is symmetric under any reordering of the possible strings is identified with a complete random source, because any other model entails biases assignations according to the strings' grouping represented by the corresponding partition. This symmetry only exists when the partition is the set $\Xi_{\beta}$ itself, hence we denote $\mathcal{M}_{\alpha^{(1)}} =\mathcal{M}_{\rm sym}
$.

Consider now that when characterising randomness the only essential feature is whether bias for or against some strings is present, but the degree of bias is irrelevant. We can eliminate the dependence on the bias parameters by multiplying with a prior for $\{\theta_r\}_{r=1}^K$ and derive the so called \textit{evidence} for a given model\cite{mackay-bayesian-interpolation}. Following\cite{myung-pnas}, we use the Jeffreys prior for it yields a model's probability distribution invariant under reparametrization and provides a measure of a model's complexity, thus giving a mathematical representation of Occam's Razor principle\cite{balasubramanian,myung-pnas,balasubramanian2}. After integrating in the parameter space, we arrive at (see  Supplementary Information (SI), Sec. 2)
\beeq{
P\left(\hat{s}|\mathcal{M}_{\alpha^{(K)}}\right)=\frac{\Gamma\left(\frac{K}{2}\right)}{\Gamma^K\left(\frac{1}{2}\right)}\prod_{r=1}^K\left(\frac{1}{|\omega^{(r)}|} \right)^{\frac{M}{\beta}\gamma_{\omega^{(r)}}} \frac{\prod_{r=1}^K\Gamma\left(\frac{1}{2}+\frac{M}{\beta}\gamma_{\omega^{(r)}}\right)}{\Gamma\left(\frac{K}{2}+\frac{M}{\beta}\right)}\,.
\label{eq:main}
}
Eq. (\ref{eq:main}) is  our main result, for it will let us perform the model selection straightforwardly. For $\mathcal{M}_{\rm sym}$, its evidence is fairly intuitive:
 \beeq{
P(\hat{s}|\mathcal{M}_{\rm sym})\equiv P\left(\hat{s}|\mathcal{M}_{\alpha^{(1)}}\right)=2^{-M}\,.
\label{eq:sm}
}

Finally, we want to infer the model that best describes our source, \emph{after} a dataset $\hat{s}$ is given. Using Bayes' theorem  the posterior distribution $P(\mathcal{M}_{\alpha^{(K)}}|\hat{s})$ reads:
\beeq{
P(\mathcal{M}_{\alpha^{(K)}}|\hat{s}) =\frac{P(\hat{s}|\mathcal{M}_{\alpha^{(K)}})P_0(\mathcal{M}_{\alpha^{(K)}}) }{\sum_\gamma P(\hat{s}|\mathcal{M}_\gamma) P_0(\mathcal{M}_\gamma)}\,.
\label{eq:feo}
}
Henceforth we will consider  a uniform prior over models (which is justified in SI),  so the model's posterior is simply proportional to its evidence. \\
Suppose now we want to assess whether a source can be considered truly random. This is performed in two steps. As the first step, we need a model ranking procedure based on the posterior distribution. The second step consists in quantifying the goodness of our choice of model.

As a decision rule for the ranking process we use the Bayes Factor\cite{robert2007-bayesian-choice} perspective,
\begin{equation}
{\rm BF}_{\alpha, {\alpha'}} = \frac{P(\mathcal{M}_\alpha|\hat{s})}{P(\mathcal{M}_{\alpha'} | \hat{s})}
 = \frac{ P(\hat{s}|\mathcal{M}_\alpha)}{ P(\hat{s}|\mathcal{M}_{\alpha'})}\,.
 \label{eq:BF}
\end{equation}
Thus, we will choose $\mathcal{M}_\alpha$ over $\mathcal{M}_{{\alpha'}}$ whenever $ {\rm BF}_{\alpha, {\alpha'}}>1$. It has been shown that $ {\rm BF}_{\alpha, {\alpha'}} $ provides a measure of goodness of fit and $\lim_{M\to \infty} {\rm BF}_{\alpha, {\alpha'}} = \infty$ if $\mathcal{M}_\alpha$ is the true model\cite{verdinelli1998-bayesian-goodness-fit}. \\
To implement the second step, which is nothing more than a hypothesis testing problem, we have two alternatives: either we check whether $\log_{10} {\rm BF}_{\alpha, {\alpha'}}\geq 2$ which is considered decisive in favour of model $\mathcal{M}_{\alpha}$ \cite{robert2007-bayesian-choice}, or we compute the ratio between the posterior and the prior of a given model to assess how certain the posterior has become under the information provided by the dataset.\\
From a computational point of view notice that the evaluation of the posterior requires to being able to compute the normalization factor $\sum_\gamma P(\hat{s}|\mathcal{M}_\gamma) P_0(\mathcal{M}_\gamma)$ that appears in (\ref{eq:feo}). When the number of models is very large we can choose either to work with a subspace of models or use the logarithm of the Bayes Factor, as in this case the normalisation factor cancels out.

It is clear that a full test of randomness requires different values of $\beta$ to be used for the same dataset, while the strings should be short enough so that the $M$ bits allow for each of the possible models to be sampled at least once. Thus, heuristically, $B_{2^{\beta_{\rm max}}}\sim M$ whence we can reproduce the BN limit\cite{Calude1993},  $\beta_{\rm max}\sim \log_2\log_2(M)$, after using an asymptotic expansion for the Bell number.

\begin{figure}
\begin{center}
\includegraphics[width=8cm,height=13cm]{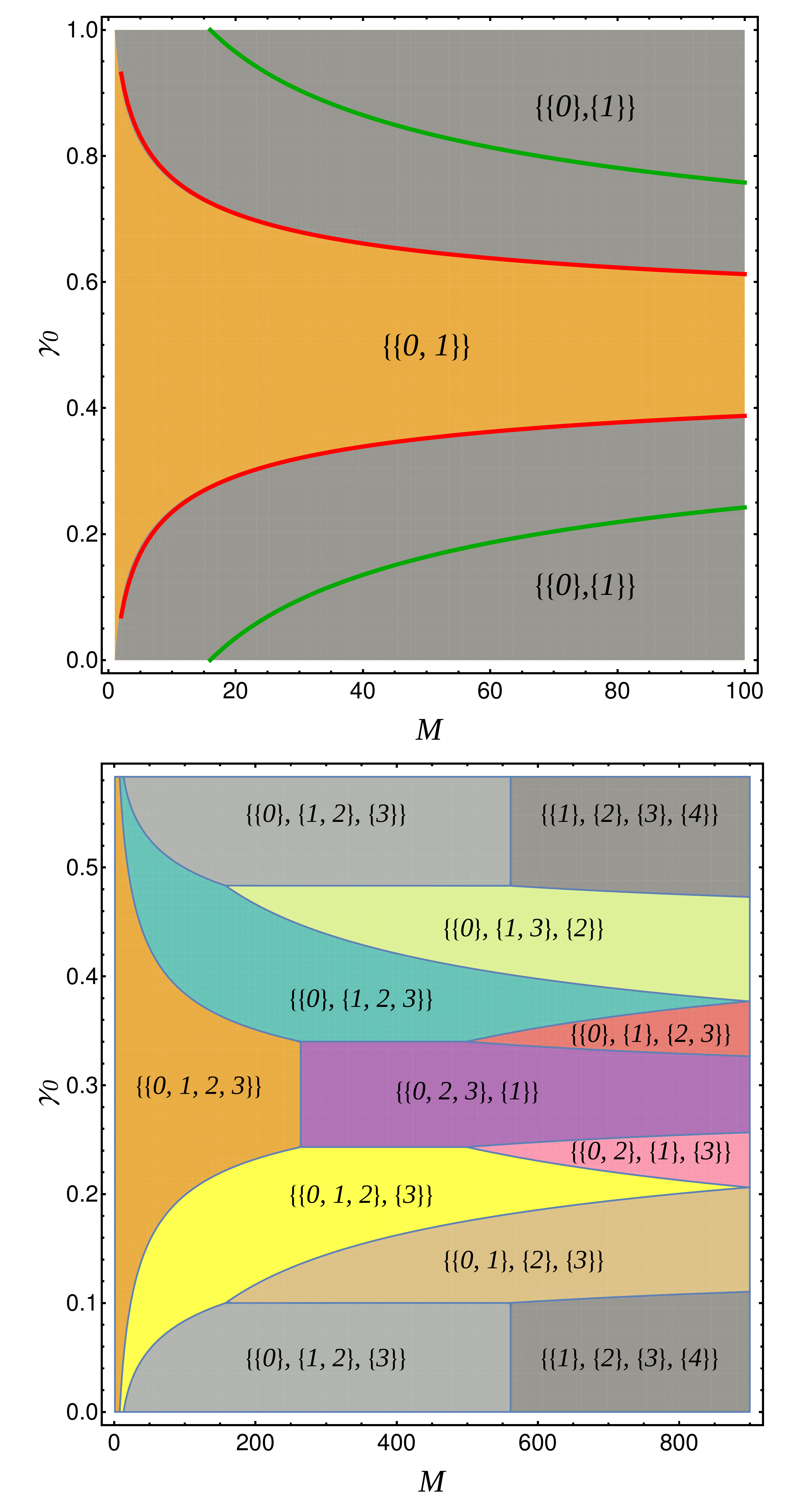}
\caption{\textbf{Phase diagram of Randomness Characterisation}. Division of the parameter space into regions according to the likeliest model. The top figure corresponds to $\beta=1$ in terms of the frequency $\gamma_0$ of the string \texttt{0} and the sample size $M$. The green curves corresponds to Borel's normality criterion, while the red curves are  Borel-type bounds obtained by an approximation obtained from  Eq. (\ref{eq:main}) (see Sec. 3 of SI). The bottom plot corresponds to $\beta=2$ where each coloured area identifies the likeliest model in that region. Here we fixed the frequencies $\gamma_1=1/6$ and $\gamma_2=1/4$ and varied the frequency $\gamma_0$ of the string \texttt{00} and the sample size $M$.
\label{fig1}}
\end{center}
\end{figure}

Note that by fixing $\beta$ we have the set of parameters  $(\{\gamma_j\}_{j=0}^{2^\beta-1},M)$, whose space can be divided into regions identifying the likeliest model according to Eq. (\ref{eq:main}). As illustrative cases, in Fig. \ref{fig1} we show a phase-type diagram for $\beta=1$ and $\beta=2$ (upper and lower panel, respectively), where the orange-filled area delimits the parameters values that renders $\mathcal{M}_{\rm sym}$ the likeliest model. The top panel includes the bounds according to the BN criterion (green curves) given by Eq. (\ref{eq:borels bound}), and shows that for any sequence length, $M$, our method allows for considerably smaller variations of $\gamma_0$. This is a significant improvement, since only necessary criteria exist for testing randomness. The lower panel depicts the analogous regions when $\beta=2$, for which there are fifteen models (see a list in the SI) and we have fixed two frequencies: $\gamma_1=1/6$ and $\gamma_2=1/4$. The complete models distribution can be deduced from the structure of this graph, by distinguishing, \textit{a posteriori}, the equiprobable strings for which the corresponding model is the likeliest. Thus more information than complete randomness classification can be readily obtained from our method.

Also in Fig.~\ref{fig1}, the red curves of the $\beta=1$ case are bounds obtained by comparing the likelihood of $\mathcal{M}_{\rm sym}$ with models involving partitions into $K=2$ subsets. Agreement with the regions boundary is excellent. Our choice of $K=2$ is justified as we would expect that models corresponding to partitions into two subsets to be the closest ones to the model $\mathcal{M}_{\rm sym}$. An explicit expression for these bounds is derived in SI, Sec. 3, and Extended Data Figures 2 and 3 depict that they also bound considerably well the region in which $ \mathcal{M}_{\rm sym} $ is the likeliest for $\beta=2$. 

\begin{figure}
\begin{center}
\includegraphics[width=8cm, height=12cm]{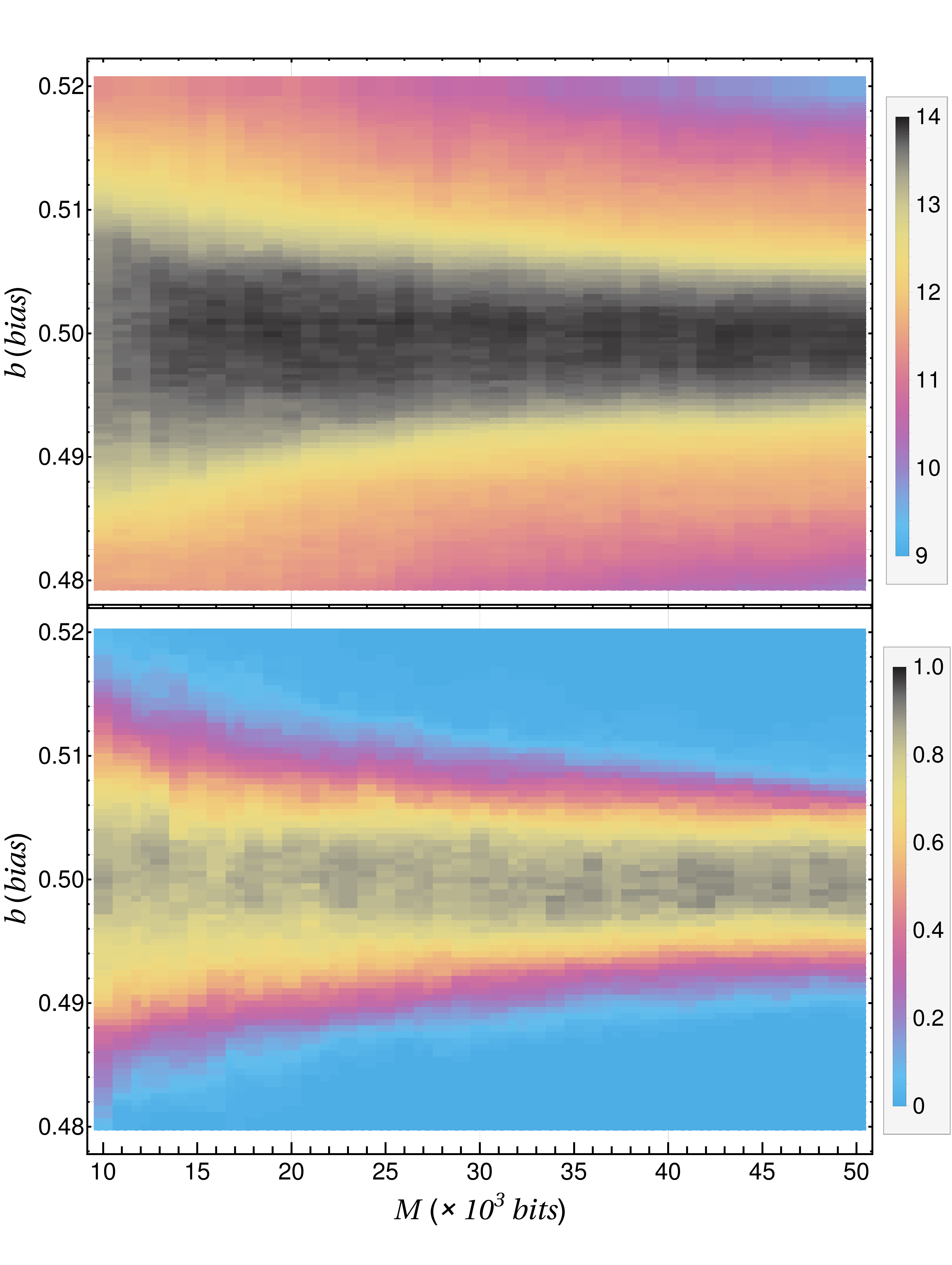}
\caption{\textbf{Comparison with NIST Suite test}. Comparison of the bias allowed on a given sequence for it to be considered random using the NIST suite (upper panel) and our Bayesian method  for randomness characterisation (lower panel). \label{fig:bayes vs nist}}
\end{center}
\end{figure}

For further benchmarking, we have compared our method against the NIST test suite\cite{NIST}. The result is depicted in Fig. \ref{fig:bayes vs nist}, as a function of the sequence length $M$ and bias $b$ employed to generate a \texttt{0}. The upper panel on Fig.~\ref{fig:bayes vs nist} shows the averaged number of tests passed when employing the NIST suite, while the lower one shows the frequency of $\mathcal{M}_{\rm sym}$ being the likeliest, for $\beta=1, 2$ and $3$. We believe that our technique can contribute to test the quality of RNG in a more stringent form, since by applying a single test thrice (once for each value of $\beta$), we determined more precisely the random character of the sample of sequences.

As an application, we have tested our method in a bit sequence obtained experimentally from the differences in time  detection in the process of spontaneous parametric down conversion (SPDC). Sequences generated via a SPDC photon-pair source  have been shown to fulfil with ease the BN criterion, and to pass comfortably the NIST's suite\cite{Solis2015}. In the SPDC process a laser pump beam illuminates a crystal with a $\chi^{(2)}$ nonlinearity, leading to the annihilation of pump photons and the emission of photon pairs, typically referred to as signal and idler\cite{burnham70}. Our experimental setup is shown in Extended Figure 1 and we explain how to construct a \texttt{0} or \texttt{1} symbol from the detection signals in Section 1 of SI. We generated a $4\times10^{9}$ bits sequence, so $\beta_{\rm max}\sim 4$. When $1\leq \beta \leq 3$, we used all the possible models in the comparison, while, for computational ease, when $\beta=4$, we restricted the model space to the $32,768$ models corresponding to $K=1$ and $K=2$ subsets (consider that $B_{2^4}=10^{10}$). Our inference showed that $\mathcal{M}_{\rm sym}$ was the likeliest model for every value of $\beta$.

As explained above, to achieve a full characterization of our QRNG as a random \emph{source}, we need to go further from the model ranking based on the Bayes Factor and measure our certainty that $\mathcal{M}_{\rm sym}$ is the true model governing the source. This (un)certainty quantification is the hallmark of Bayesian statistics, since $P(\mathcal{M}_{\rm sym}|\hat{s})$ represents the probability that modelling our QRNG as a random source is correct. Computing this posterior distribution directly from Bayes' Theorem, Eq. \ref{eq:feo}, we arrive at the values shown in Table \ref{table1} for each $\beta$. The first three values are at least $0.95$, but the corresponding to $\beta=4 $ is about $0.32 $, considerably smaller. However, this represents an improvement of order $10^4$ when compared with the initial value for the prior, $P_0(\mathcal{M}_{\rm sym}) = 1/{32,768} \approx 3.1\times 10^{-5}$. Alternatively, we computed  $\log_{10} {\rm BF}_{{\rm sym},\alpha'}$ for each value of $\beta$. The values reported in Table \ref{table1} correspond to the comparison of $\mathcal{M}_{\rm sym}$ and the second likeliest model, hence the inequality for $\beta >2$. These two criteria combined lead us to conclude that there is decisive evidence for our hypothesis that $\mathcal{M}_{\rm sym}$ is the underlying model driving our source, thus verifying that the photonic RNG is strictly random in the sense described in the article.

\begin{table*}[h]
\caption{Posterior $P(\mathcal{M}_{{\rm sym}}|\hat{s})$ calculated for a dataset of $4\times 10^9$ bits.}
\begin{center}
\begin{tabular}{@{\vrule height 10.5pt depth4pt  width0pt} c c c }

\vrule depth 6pt width 0pt $\beta$ & $P(\mathcal{M}_{{\rm sym}}|\hat{s})$ &$\log_{10} {\rm BF}_{{\rm sym},\alpha'}$ \\
\hline
1 & 0.999965 & $ 4.45$ \\
2 & 0.999562  & $ \geq 3.72 $\\
3 & 0.968353  & $ \geq 2.01$\\
4 & 0.46718 &$\geq3.46 $\\
\hline
\end{tabular}
\end{center}
\label{table1}
\end{table*}

From a more general perspective, we propose that $P(\mathcal{M}_{\alpha^{(K)}}|\hat{s})$ quantifies our certainty on the hypothesis that a sequence $\hat{s}$ was generated using the biases on strings associated with $\alpha^{(K)}$. Because Bayesian methods entails a model's generalizability\cite{mackay-bayesian-interpolation,myung-pnas}, the likeliest model provides a characterization of the source of $\hat{s}$. All partitions can be identified with standard computational packages, although it can be computationally demanding for sequences of $\sim 10^{10}$ bits. In any case, once a partition is given, its model's likelihood is easily found using Eq. (\ref{eq:main}). A simplified analysis can be performed with the BN-type bounds given in Section 3 of the SI, which also leads to more stringent criteria than other approaches.

\begin{addendum}
\item[Supplementary Information] is linked to the online version of the paper at www.nature.com/nature
\item IPC and RDHR thank hospitality to the Abdus Salam ICTP. RDHR also  thanks  Susanne Still, Valerio Volpati, and Aaron King for helpful discussions regarding the choice of models priors. This work has received partial economical support from Consejo Nacional de Ciencia y Tecnolog\'ia (Conacyt): SEP-Conacyt and RedTC-Conacyt, Mexico, PAPIIT-UNAM project IN109417, and PAPIIT-UNAM project IA103417. We also want to thank Mark Buchanan for his helpful feedback for writing the manuscript.
\item[Author Contributions] I.P.C., R.D.H.R. and M.M. developed the Bayesian approach for the current application and derived the analytic expressions for the evidence of models. A.S., J.G.H.. A.U., and A.M.A.M. furnished our work as a randomness characterization and provided the experimental datasets. The comparison with the NIST test suite and BN criterion was done by R.D.H.R. and A.S. All authors discussed the results and commented the manuscript.
 \item[Competing Interests] The authors declare that they have no competing financial interests.
 \item[Correspondence] Correspondence and requests for materials should be addressed to Isaac P\'erez Castillo.~(email: isaacpc@fisica.unam.mx).
\end{addendum}

\pagebreak

\renewcommand{\figurename}{Extended Data Figure}
\setcounter{figure}{0}

\section*{\huge Supplementary Information}

\section{Experimental Setup and conversion to a sequence of random bits.}

The quantum state of the emitted photon pairs can be written as $|\Psi\rangle=|\mbox{vac}\rangle+\eta |\Psi_2\rangle$ in terms of the vacuum $|\mbox{vac}\rangle$, the two-photon component $|\Psi_2\rangle$, and of a constant $\eta$ related to the conversion efficiency.   Under the assumptions a continuous-wave, plane-wave pump $ |\Psi_2\rangle$ may be expressed as\cite{Vicent2010} 
\begin{equation}
\label{E:state}
 |\Psi_2\rangle=\int d \omega \int d \textbf{k}^\bot F(\omega,\textbf{k}^\bot) |\omega,\textbf{k}^\bot  \rangle_s |\omega_p-\omega,-\textbf{k}^\bot  \rangle_i,
\end{equation}
\noindent written in terms of a joint amplitude function $F(\omega,\textbf{k}^\bot)$, and  where $|\omega,\textbf{k}^\bot  \rangle_\mu$ represents a single-photon Fock state with frequency $\omega$ and transverse wavevector $\textbf{k}^\bot$ for mode $\mu$, with $\mu=s,i$ for the signal ($s$) and idler ($i$). In writing the two-photon state, we have assumed that the parametric downconversion process is in the  spontaneous regime, so that the appearance of multiple-pair events can be neglected. This assumption is valid if the parametric gain is sufficiently low; experimentally, we restrict the pump power so that the process remains spontaneous.  In all likelihood, a similar experiment and analysis carried out in the high-gain, stimulated regime would yield different results from those presented on this paper. 

The state in Eq. (\ref{E:state}) is entangled since it cannot be factored into a direct product of separate states $|S\rangle $  (signal) and $|I\rangle$ (idler) as $|\Psi\rangle=|S\rangle |I\rangle$.  While in many works based on SPDC photon pairs entanglement is the key resource, in our case we  exploit instead the random times of emission (and detection) of signal and idler photons.

\begin{figure*}
\begin{center}
\includegraphics[width=8cm]{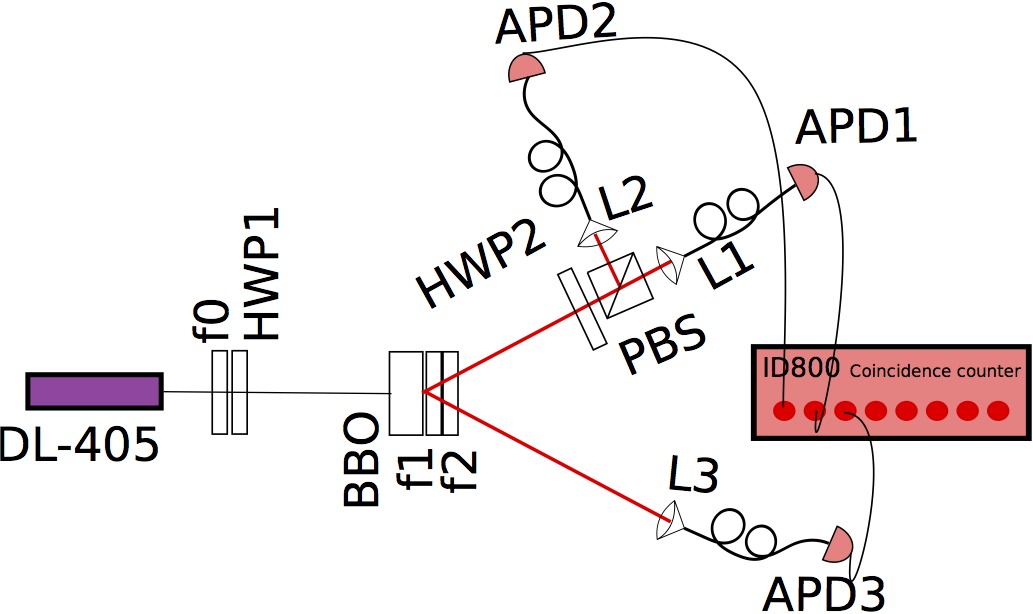}
\caption{\textbf{Experimental Setup}. A pump laser beam centred at 407nm (DL407) incides into nonlinear BBO crystal. The signal and idler generated photons are emitted at diametrically opposed portions of an emission cone which yields phase matching for frequency-degenerate non-collinear photon pairs. A polarising beam splitter (PBS) and a Half wavelength plate (HWP2) are placed at the signal portion of the cone so this photon can be transmitted or reflected with a 50/50 probability, the reflected and transmitted signal and idler photons are collected into multimode fibers that lead to avalanche photodiodes(APD1,2,3) which emit a TTL pulse for each detection event.}
 \label{ext-fig:expsetup}
\end{center}
\end{figure*}

We have used a pump beam from a diode laser (DL407) centred at $407$nm with $\sim60$mW power, and as nonlinear medium a  $\beta$ barium borate (BBO) crystal of $1$mm length; see  Extended Data Figure \ref{ext-fig:expsetup}.   The BBO crystal, which is negative uniaxial,  was cut so that the angle subtended by the optic axis with respect pump beam axis is $\theta_{\mbox{pm}}=29.2^\circ$ which yields phase matching for the generation of frequency-degenerate, non-collinear photon pairs.  Signal and idler photons are emitted on diametrically opposed portions of an emission cone centred on the pump beam axis,  with a $3.6^\circ$ half opening angle.  Pump photons are suppressed by transmitting the signal and idler modes through a long-pass filter which transmits wavelengths  $\lambda>488$nm (F1), followed by a bandpass filter centred at $800$nm with a $40$nm bandwidth (F2). 

A  halfwaveplate (HWP2) and a polarising beam splitter (PBS) are placed on the signal arm so that the signal photon is transmitted or reflected with 50/50 probability.  Each of the idler, reflected signal and transmitted signal collection modes is defined by an $f=8$mm focal length aspheric lens  (L1, L2 and L3) which focuses incoming light into the core of a multi-mode fibre with a $50\mu$m diameter (MMF1, MMF2 and MMF3).  The plane defined by the collection fibres is chosen for convenience to be parallel to the optical table.   By monitoring coincidences between the reflected signal and idler modes, on the one hand, and between the transmitted signal and idler modes, on the other hand, we are able to probabilistically exclude double (and multiple) pair events.

Each of the three photon-collection fibres leads to a silicon-based avalanche photodiode (APD1, APD2 and APD3), which emits an electronic TTL pulse for each detection event.  The times of arrival of these pulses are monitored with a time to digital converter (TDC; id800 from IdQuantique), or time-tagger, with a resolution of 81 ps. The TDC produces three time series containing the time of arrival data for each of the idler ($i_n$), and  transmitted ($s^t_n$) and reflected ($s^r_n$) signal channels.  We generate by post-processing the two time series defined as $c^t_n=s^t_n \times i_n$, and $c^r_n=s^r_n \times i_n$,  
corresponding to those bins for which there are coincident detection events between the (reflected or transmitted) signal and idler channels. A sequence of bits is generated by comparing the differences in time  detection with a fully regular time series \emph{with the same number of events per second}.  A value of $1$ is assigned if the time of detection is smaller than the corresponding time in the regular time series, and a value of $0$ otherwise\cite{Solis2015}.\\

We have checked  on the efficiency of our QRNG in our experimental setup. According to our data, the efficiency based on the SPDC is 240 kilocounts per second in each channel.  If only those events in which the signal and the idler photon are detected in coincidence are registered, the efficiency of random number generation is reduced to 27 kilocounts per second. Moreover, our experimental setup is such that we are able  to discriminate  four-photon versus two-photon events. This is achieved by noticing that, first of all, we have used a pump power such that the rate of four-photon generation is essentially negligible: less than 0.2\% according to our data. Secondly, in one of the SPDC arms we  have placed a beamsplitter so that by discarding those events in which both APD's in that arm click, we can eliminate all the events in which events are detected in same time bin in the three detectors.

\section{Derivation of Jeffreys Prior and Model's evidence}
The idea of the Jeffreys prior is to take into account model indistinguishability from a point of view of a statistical sample. Based on Sanov's theorem\cite{Mastromatteo2013} we know that the volume of models which are indistinguishable is inversely proportional to the square root of the determinant of the Fisher information matrix. This idea of measuring relevant volumes across models, but using a graining approach has also been explored previously\cite{myung-pnas,balasubramanian}
in a rigorous geometric treatment. Note that in this case, our parameters are the $\theta$'s of which only the (say) first $K-1$ are independent due to the normalization requirement. Then, considering a model $\mathcal{M}_{\alpha^{(K)}}$ -- also obviating the index $\ell$ in the partition, as we did in the main text -- we have the following minus log-likelihood for a string $s$ assigned by $\mathcal{M}_{\alpha^{(K)}}$ to partition $\omega^{(a)}$
\begin{eqnarray*}
- \log P\left(s|\mathcal{M}_{\alpha^{(K)}},\{\theta_r\}\right)=
- \log\left(\frac{\theta_a}{|\omega^{(a)}|}\right)
\end{eqnarray*}
From here we derive the Fisher information matrix $J_{ab}$ for $a,b=1,\ldots,K$
\begin{eqnarray*}
J_{ab}(\theta)=-
{\rm E}\left[\frac{\partial^2}{\partial \theta_a\partial \theta_b}\log P\left(s|\mathcal{M}_{\alpha^{(K)}},\{\theta_r\}\right)\right] \propto
\frac{1}{\theta_a}\delta_{a,b} \,, \nonumber
\end{eqnarray*}
where $\rm E[\cdots]$ denotes the expected value. Its determinant is  simply $\det [J_{ab}(\theta)] \propto \frac{1}{\prod_{r=1}^K\theta_r}$. The proportionality constants will cancel out, once we normalize our expression for $P_{\rm{Jeff}}$.
From here we have the following expression for Jeffreys prior:
\begin{eqnarray}
P_{\rm Jef}(\theta)=\frac{\Gamma\left(\frac{K}{2}\right)}{\Gamma^K\left(\frac{1}{2}\right)}\prod_{r=1}^K\theta^{-1/2}_r\,,
\end{eqnarray}
where the normalization factor comes from:
\begin{eqnarray*}
\int \left[\prod_{r=1}^K d\theta_r\right]\left[\prod_{r=1}^K\theta^{-1/2}_r\right]\delta\left(\sum_{r=1}^K \theta_r-1\right)=\frac{\Gamma^K\left(\frac{1}{2}\right)}{\Gamma\left(\frac{K}{2}\right)}\,.
\end{eqnarray*}
Notice that in this case the Jeffreys prior always behaves as a proper one, that is, it is normalizable.

Finally, a similar integration shows that the model's evidence is given by\cite{mackay-bayesian-interpolation}
\begin{eqnarray}
P\left(\hat{s}|\mathcal{M}_{\alpha^{(K)}}\right)&=&
\int \left[\prod_{r=1}^K d\theta_r\right]P_{\rm Jef}(\theta)P\left(\hat{s}|\mathcal{M}_{\alpha^{(K)}},\{\theta_r\}\right) \nonumber \\
&=&
\frac{\Gamma\left(\frac{K}{2}\right)}{\Gamma^K\left(\frac{1}{2}\right)}\prod_{r=1}^K\left(\frac{1}{|\omega^{(r)}|}\right)^{k_{\omega^{(r)}}}\frac{\prod_{r=1}^K\Gamma\left(\frac{1}{2}+k_{\omega^{(r)}}\right)}{\Gamma\left(\frac{K}{2}+\frac{M}{\beta}\right)}\,.
\label{eq:a}
\end{eqnarray}
This allows us to identify the terms $ \left(\frac{1}{|\omega^{(r)}|}\right)^{k_{\omega^{(r)}}} $ as the maximum likelihood estimators, and the ones involving the gamma functions as a measure of the relevant volume occupied in the parameter space, related to the model's complexity\cite{myung-pnas}.

\section{Borel-normality-type (BN-type) bounds}

Suppose we are interested in discerning whether a given sequence is completely random or not. This means that we must look for the region in the parameter space $(\{\gamma_j\}_{j\in\Xi_\beta},M)$ in which the evidence of the symmetric model --corresponding to the partition of $\Xi_\beta$ into one subset-- is bigger than the rest of the models. As the empirical frequencies $\{\gamma_j\}_{j\in\Xi_\beta}$ are grouped into $K$  subsets for a given partition $\alpha^{(K)}$, then the corresponding model has in effect $K-1$ free parameters $\{\gamma_{\omega^{(r)}}\}_{r=2}^{K}$. Recalling  that we used the Bayes Factor as a decision rule in the main text, we can explore the conditions such that $\mathcal{M}_{\rm sym}$ is the likeliest by the behaviour of the log-likelihood ratio, $\log\left(\frac{P(\hat{s}|\mathcal{M}_{\rm sym})}{P(\hat{s}|\mathcal{M}_{\alpha^{(K)}})}\right)$.

To obtain a BN-type bound, we do the following: i) look for the values  $\{\gamma^\star_{\omega^{(r)}}\}_{r=2}^{K}$  which extremize the log-likelihood ratio; ii) do an expansion around those values up to second order. We eventually obtain:
\begin{eqnarray}
&&\log\left(\frac{ \prod_{r=1}^K\left(|\omega^{(r)}| \right)^{\frac{M}{\beta}\gamma^{\star}_{\omega^{(r)}}}\Gamma^K\left(\frac{1}{2}\right)\Gamma\left(\frac{K}{2}+\frac{M}{\beta}\right)}{2^{M}\Gamma\left(\frac{K}{2}\right)\prod_{r=1}^K\Gamma\left(\frac{1}{2}+\frac{M}{\beta}\gamma^{\star}_{\omega^{(r)}}\right)}\right)\nonumber\\
&&=\frac{1}{2}\left(\frac{M}{\beta}\right)^2\sum_{r,r'=2}^K\left(\gamma_{\omega^{(r)}}-\gamma^{\star}_{\omega^{(r)}}\right)\left(\gamma_{\omega^{(r')}}-\gamma^{\star}_{\omega^{(r')}}\right)\nonumber\\
&&\times\left[\delta_{r,r'}\psi_1\left(\frac{1}{2}+\frac{M}{\beta}\gamma^{\star}_{\omega^{(r)}}\right)+\psi_{1}\left(\frac{1}{2}+\frac{M}{\beta}\left(1-\sum_{r=2}^K \gamma^{\star}_{\omega^{(r)}}\right)\right)\right]\,,
\label{eq:ctbsb}
\end{eqnarray}
where the $\gamma^\star$-unknowns obey the following set of equations
\beeq{
\psi\left(\frac{1}{2}+\frac{M}{\beta}\gamma^{\star}_{\omega^{(r)}}\right)-\psi\left(\frac{1}{2}+\frac{M}{\beta}\left(1-\sum_{r=2}^K \gamma^{\star}_{\omega^{(r)}}\right)\right)\\
=\log \left|\frac{\omega^{(r)}}{\omega^{(1)}}\right|\,,\quad\quad r=2,\ldots,K\,.
\label{eq:ctbsb3}
}
Here the function $\psi_n(x)$ is the polygamma function of order $n$, with $\psi(x)\equiv \psi_0(x)$. As the symmetric model is the one that corresponds to no-free parameters, one could reasonable assume that the models which are closer to $\mathcal{M}_{\rm sym}$ are those which correspond a single free parameter. This, in turn, corresponds to subfamilies of partitions into two subsets of lengths $\{2^{\beta}-q,q\}$ for $q=1,\ldots, 2^{\beta}/2$, which will have aggregate frequencies $1-\gamma_{|q|}$ and $\gamma_{|q|}$ respectively. This is also justified by the lower panel of Figure 1 in the main text, which shows that the transition from $K = 1$ to a bigger value should necessarily go through a region where a model with $K = 2$ is likelier than $\mathcal{M}_{\rm sym}$. Applying this to the set of Eqs. (\ref{eq:ctbsb}) and  (\ref{eq:ctbsb3}) we obtained that $\left|\gamma_q-\gamma^\star_{|q|}\right|\leq\frac{\sqrt{2}\beta}{M} \mathcal{W}(\gamma^{\star}_{|q|})$  with the function $\mathcal{W}(\gamma^{\star}_{|q|})$ defined as
\beeq{
\mathcal{W}(\gamma^{\star}_{|q|})\equiv    \sqrt{\frac{\log\left(\frac{\Gamma^2(1/2) \Gamma(1+M/\beta)\left(2^\beta-q \right)^{\frac{M}{\beta}(1-\gamma^\star_{|q|})}q^{\frac{M}{\beta}\gamma^{\star}_{|q|}}}{2^{M} \Gamma\left(\frac12+\frac{M}{\beta}\gamma^{\star}_{|q|}\right)\Gamma\left(\frac12+\frac{M}{\beta}(1-\gamma^{\star}_{|q|})\right)}\right)}{\psi_1\left(\frac12+\frac{M}{\beta}\gamma^{\star}_{|q|}\right)+\psi_{1}\left(\frac12+\frac{M}{\beta}\left(1-\gamma^{\star}_{|q|}\right)\right)}}\,,
\label{eq:ctbsc}
} 
where  $\gamma^{\star}_{|q|}$ are the aggregated frequencies of a subset of size $q$ satisfying the extremisation condition
\beeq{
\psi\left(\frac12+\frac{M}{\beta}\gamma^{\star}_{|q|}\right)-\psi\left(\frac12+\frac{M}{\beta}\left(1- \gamma^{\star}_{|q|}\right)\right)=\log \left(\frac{q}{2^{\beta}-q}\right)\,,
\label{eq:ctbsc2}
}
for $q=1,\ldots,2^{\beta}/2$.

\begin{figure*}
\begin{center} 
\includegraphics[width=\textwidth, height=6cm]{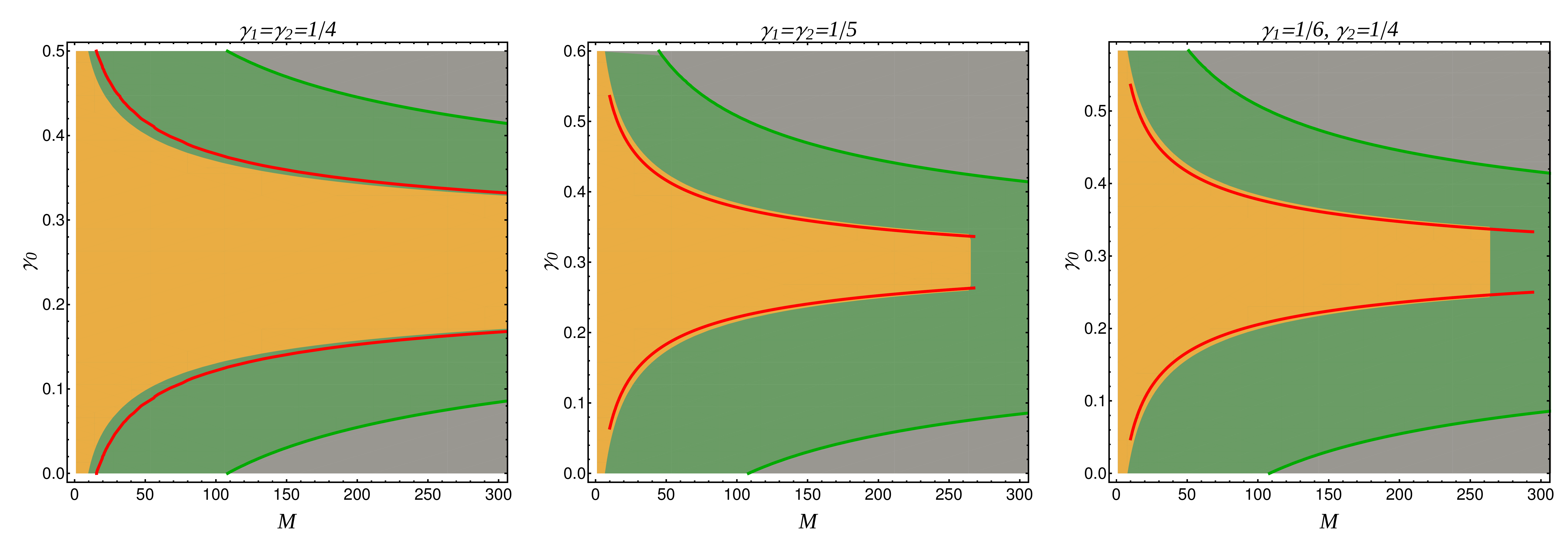}
\caption{\textbf{BN-type bounds}. Phase diagram of model selection for the 15 models for $\beta=2$ and various fixed values of $\gamma_1$ and $\gamma_2$. Here the orange filled  area represents the region in which  model $\mathcal{M}_{\rm sym}$ is the likeliest, while the grey filled area represents the region in the parameter space in which any other model is the likeliest. Solid red lines represent the BN-type bounds.  We also compare with the BN bounds (green filled region). Notice that for the second and the third case, the BN bounds also provides a bound for $M$ given by the solution of $|1/4-1/5|=\sqrt{\frac{\log_2(M)}{M}}$ and $|1/4-1/6|=\sqrt{\frac{\log_2(M)}{M}}$, respectively. }
\label{ext-fig:ctbbeta2}
\end{center}
\end{figure*}

In particular, for $\beta=1$, there is only one model to compare to $\mathcal{M}_{\rm sym}$, which  precisely  corresponds to $K=2$. Here, the solution of (\ref{eq:ctbsc2}) is exactly $\gamma_{|1|}^{\star}=1/2$, which provides the following bound:
\beeq{
\left|\gamma_1-\frac{1}{2}\right|\leq \frac{1}{M}\sqrt{\frac{\log\left(\frac{2^{-M} \Gamma(1+M)}{\Gamma^2\left(\frac{1}{2}+\frac{M}{2}\right)}\right)}{\psi_1\left(\frac{1}{2}+\frac{M}{2}\right)}}\,,
\label{eq:ctbbeta1}
}
This is the formula we used to draw the red curves in the top panel of Figure 1 of the main text together with the exact diagram. Agreement for this simple bound is excellent compared to the exact formulas, and rather different as compared to the one of BN. For the case $\beta=2$, one must solve the set of equations numerically to evaluate the bounds. They work reasonably well in the parameter space and much better than the BN bounds as shown in Extended Data Figure \ref{ext-fig:ctbbeta2}. Notice that in these figures we only depict two regions in the parameter space: the orange one corresponds to the region in which the symmetric model is likeliest, while the grey-filled area in which it is not.

\begin{figure*}
\begin{center}
\includegraphics[width=\textwidth, height=6cm]{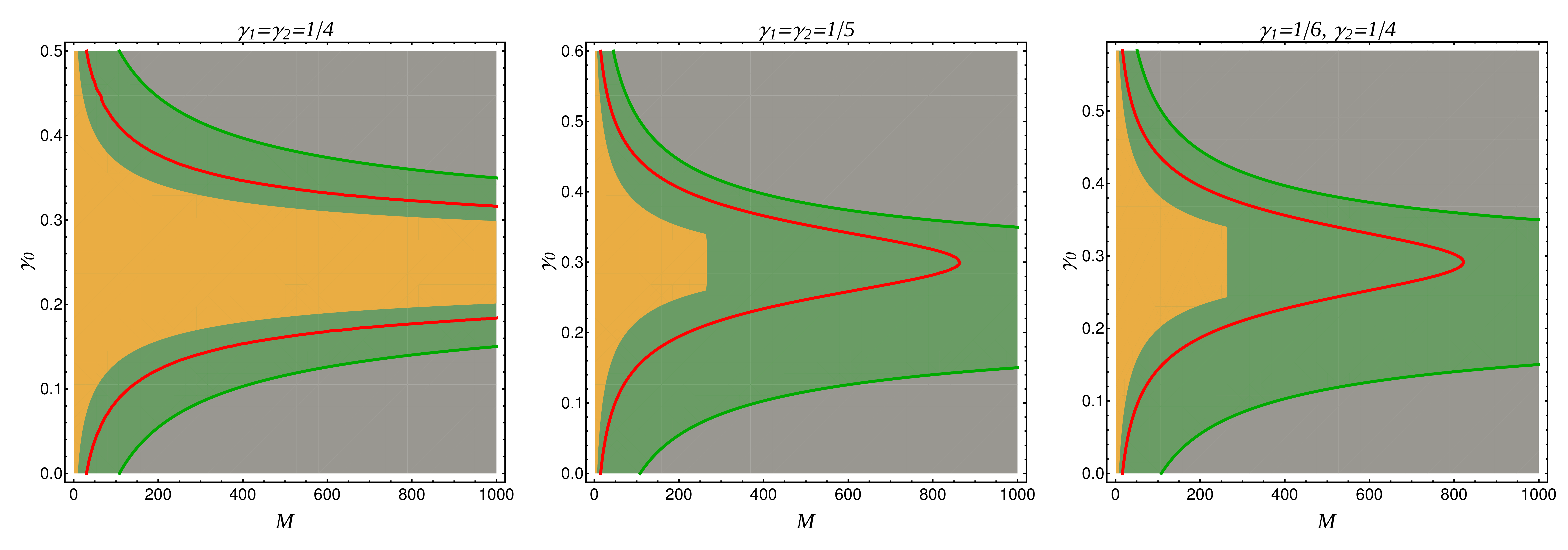}
\caption{\textbf{BN-type bounds}. Phase-type diagram for model selection for $\beta=2$ and comparison between the bounds given by the simple formula (\ref{eq:ctbsc}) (solid red line) and the Borel-normality bounds (solid green line).}
\label{ext-fig4}
\end{center}
\end{figure*}

These previous bounds have the disadvantage of needing to solve the system (\ref{eq:ctbsc2}) numerically. However, looking at the set of Eqs. (\ref{eq:ctbsb3}) we notice that there is a particular set of partitions for which its solution is particularly simple, namely when the system is solved using only equi-partitions, that is, partitions into subsets of the same size. With this restriction, it is possible to find simpler, less restrictive bounds, yet tighter than the ones derived from other methods.  Suppose that we look at partitions into $K$ subsets. Within this family (and of course for even $K$) we will have a subfamily of equi-partitions. For them we have that $|\omega^{(r)}|={|\omega^{(1)}|}=\frac{2^{\beta}}{K}$ and therefore $k^{\star}_{\omega^{(r)}}=M/(\beta K)$ and $\gamma^{\star}_{\omega^{(r)}}=1/K$.  In particular, for the  model corresponding to  a partition into $K=2^\beta$ subsets,  the formula (\ref{eq:ctbsb}) becomes:
\beeq{
\sum_{i\leq j=1}^{2^{\beta}-1}\left(\gamma_{i}-\frac{1}{2^{\beta}}\right)\left(\gamma_{j}-\frac{1}{2^{\beta}}\right)=
\left(\frac{\beta^2\log\left(\frac{2^{-M} \Gamma^{2^\beta}(\frac12)\Gamma\left(2^{\beta-1} +\frac{M}{\beta}\right)}{\Gamma\left(2^{\beta-1} \right) \Gamma^{2^\beta}\left(\frac{1}{2}+\frac{M}{\beta 2^\beta}\right)}\right)}{M^2\psi_1\left(\frac{1}{2}+\frac{M}{\beta}\right)}\right)
\label{eq:ctbs4}\,,
} 
a bound which, unlike the one of Borel-normality, couples all the empirical frequencies. Results of these broader bounds are plotted in Extended Data Figure \ref{ext-fig4}. 

\section{Some examples for the evidence}
In this section, we illustrate, with some specific examples, the formulae Eq. (4) for the particular case of $\beta=2$. Because explicit reference to specific partitions is made, we will use the full notation $\alpha^{(K)}_\ell$, although there is no natural order to assign the index $\ell$. In this case we have the following partitions of $\Xi_{\beta=2}$, corresponding to 15 models: a partition  into $K=1$ subset (symmetric model) which corresponds to $\alpha^{(1)}_1=\{\{0,1,2,3\}\}$. There are  $\left\{ {4 \atop 2} \right\}=7$ ($\left\{ {a \atop b} \right\}$ denotes the Stirling number of second kind)  partitions  with $K=2$ subsets, which are: $\alpha_1^{(2)}=\{\{0\}, \{1, 2, 3\}\}$, $\alpha_2^{(2)}=\{\{0, 1\}, \{2, 3\}\}$, $ \alpha_3^{(2)}=\{\{0, 2, 3\}, \{1\}\}$, $\alpha_4^{(2)}= \{\{0, 1, 2\}, \{3\}\}$, $\alpha_5^{(2)}=\{\{0, 3\}, \{1, 2\}\}$, $\alpha_6^{(2)}=\{\{0, 1, 3\}, \{2\}\}$, $\alpha_7^{(2)}=\{\{0, 2\}, \{1, 3\}\}$.  We have $\left\{ {4 \atop 3} \right\}=6$  partitions into $K=3$ subsets: $
\alpha_1^{(3)} =\{\{0\}, \{1\}, \{2, 3\}\}$, $\alpha_2^{(3)}= \{\{0\}, \{1, 2\}, \{3\}\}$, $\alpha_3^{(3)}=  \{\{0\}, \{1, 3\}, \{2\}\}$, 
$\alpha_4^{(3)} =   \{\{0, 1\}, \{2\}, \{3\}\}$, $\alpha_5^{(3)}=   \{\{0, 2\}, \{1\}, \{3\}\}$,  $\alpha_6^{(3)}=     \{\{0, 3\}, \{1\}, \{2\}\}$. And, finally, one partition $\alpha_1^{(4)}=\{\{0\},\{1\},\{2\},\{3\}\}$ into $K=4$ subsets.  

An example of the evidence, of the model associated to partition e.g. $\alpha_{1}^{(3)}$ is
\begin{eqnarray}
P\left(\hat{s}|\mathcal{M}_{\alpha_{1}^{(3)}}\right) &=& 
\frac{\Gamma\left(\frac{3}{2}\right)}{\Gamma^3\left(\frac{1}{2}\right)}\left(\frac{1}{2} \right)^{k_{\omega^{(3)}}}\frac{\Gamma\left(\frac{1}{2}+k_{\omega^{(1)}}\right)\Gamma\left(\frac{1}{2}+k_{\omega^{(2)}}\right)\Gamma\left(\frac{1}{2}+k_{\omega^{(3)}}\right)}{\Gamma\left(\frac{3}{2}+\frac{M}{2}\right)} \\[4mm]
&=&
\frac{\Gamma\left(\frac{3}{2}\right)}{\Gamma^3\left(\frac{1}{2}\right)}\left(\frac{1}{2} \right)^{k_{2}+k_3}\frac{\Gamma\left(\frac{1}{2}+k_{0}\right)\Gamma\left(\frac{1}{2}+k_{1}\right)\Gamma\left(\frac{1}{2}+k_{2}+k_3\right)}{\Gamma\left(\frac{3}{2}+\frac{M}{2}\right)}\,,
\end{eqnarray}
where $ k_{\omega^{(1)}} $ ($ k_{\omega^{(2)}} $)  is the number of occurrences of string $\{0\} = \{00\}$ (resp. $\{1\} = \{01\}$), and $k_{\omega^{(3)}} $ is the added number of occurrences of the strings $\{2\} = \{10\}$ and $\{3\} = \{11\}$ in the sequence of bits. An equivalent expression with the individual frequencies $k_j$ of the $j$-th string is also given for clarity.

\section{On the choice for the Prior of models}
Since in this work our particular goal is to assess the randomness of a given sequence with a general applicable method, it would be convenient to obtain a criterion as sharp as possible when no previous knowledge of the source producing the data is given. Morevover, another desirable property would be  that no particular type of sequence is preferred over the rest, or in other words, we would like to reproduce a distribution on datasets that resembles closely a uniform prior distribution over them. As we will justify here, those two features can be achieved by choosing  a uniform prior distribution on the models, that is, for a fixed $\beta$, $P_0(\mathcal{M}_\alpha) = \frac{1}{B_{2^\beta}}$, with $B_n$ the $n$-th Bell number. Indeed, this results in a distribution on sequences for which the unbiased ones are the most unlikely.

Indeed, first of all, we need to relate the  prior distribution on models $P_0(\mathcal{M}_\alpha)$ with the prior distribution on sequences $P_0(\hat{s}) $. This can be done by computing the marginal of their joint distribution, $P_0(\hat{s}) = \sum_\alpha P(\hat{s}|\mathcal{M}_\alpha) P_0(\mathcal{M}_\alpha)$. We want to show that a uniform prior on models results into an expression of  $P_0(\hat{s})$ that penalizes unbiased sequences. To be specific, let us analyse the case of $\beta=1$, for which there are only two possible models, and hence $P_0(\mathcal{M}_\alpha)=\frac{1}{2}$. Using Eqs.~(4) and (5) from the main text to calculate the above marginal, we obtain the following
\begin{equation}
P_0(\hat{s})  = \frac{1}{2} \left[  \frac{1}{2^M} + \frac{\Gamma(1/2) \Gamma\left(k_0+1/2\right ) \Gamma\left(k_1+1/2\right ) }{ \Gamma\left(M+1\right )  \Gamma^2(1/2) }  \right]\,. 
\end{equation}
From this expression, we can see that under the assumption of uniform prior  distributions over \emph{models}, we obtain two terms for the prior distribution on \emph{datasets}: the first one is independent on the frequency of strings, while the second  term  adds a non-negative contribution that depends explicitly on such frequencies. However, this second term is just the $B$ function, whose global minimum is achieved when $k_0=k_1=M/2$. Thus unbiased sequences for which presumably $k_0 \approx k_1$ are unfavored with this assumption. 

An  analogous argument follows straightforwarldy for larger values of $\beta$. It is also worth mentioning that were we to assume directly that $P_0(\hat{s})=\frac{1}{2^M}$, the only compatible prior over models would be $P_0(\mathcal{M}_\alpha) = \delta_{{\rm sym},\alpha}$.

\end{document}